\documentclass[multphys,vecphys]{svmult}

\usepackage{makeidx}     
\usepackage{graphicx}    
                         
\usepackage{multicol}    
\usepackage{threeparttable}

\makeindex             
                       
\begin{document}

\title*{High-Energy Radiation Probes of Protostellar Envelopes}
\author{P. St\"auber\inst{1}\and
A.O. Benz\inst{1}\and
S.D. Doty\inst{2}\and
E.F. van Dishoeck\inst{3}\and
J.K. J{\o}rgensen\inst{3}}
\institute{Institute of Astronomy, ETH-Zentrum, CH-8092 Zurich, Switzerland
\texttt{pascalst@astro.phys.ethz.ch} \texttt{benz@astro.phys.ethz.ch}
\and Department of Physics and Astronomy, Denison University, Granville, OH 43023, USA
\texttt{doty@cc.denison.edu}
\and Sterrewacht Leiden, PO Box 9513, 2300 RA Leiden, The Netherlands
\texttt{ewine@strw.leidenuniv.nl} \texttt{joergens@strw.leidenuniv.nl}}
\maketitle

We present observations of molecular high-energy radiation probes and report the first detections of CO$^+$ toward W$3$ IRS$5$, a source containing ultracompact HII regions. UV radiation and X-rays from the central objects may enhance molecules due to photodissociation and ionization processes. To study the effects on the immediate YSO environment, we are developing time- and depth-dependant chemical models containing UV and X-ray chemistry, by extending the models of Doty et al. ($2002$). Molecules like CO$^+$ or NO may be used as tracers of very early X-ray emission in regions of high extinction, from which no X-ray photons can be detected.

\section{Introduction}
\label{sec:1}
In the very earliest stage of star formation, the young stellar object is surrounded by a collapsing envelope containing mainly dust and a large variety of molecules. Recently modeling of the dust continuum emission from the envelopes around high-mass (van der Tak et al. $1999$, $2000$) and low-mass protostars (J{\o}rgensen et al. $2002$) has been used to establish their physical parameters and as input for radiative transfer modeling in order to establish their chemical structure. The chemistry in these envelopes is complex and has several chemical characteristics with different diagnostics (van Dishoeck $2002$). Although molecular clouds exposed to external UV and X-rays have been studied for decades, very little is known about the influence of central UV radiation and X-rays on the chemical evolution of protostellar envelopes.

In recent years it has become clear that YSOs are strong X-ray emitters. The observed X-ray luminosities range from approximately $10^{28.5}$ to $10^{33}$\,erg\,s$^{-1}$ (Feigelson \& Montmerle $1999$, Preibisch $1998$). The emission may originate from accretion, from magnetic activity or, in massive stars, from stellar wind instabilities. Since the emission starts in the deeply embedded phase, its onset and initial intensity cannot be observed directly due to very high extinction. It is therefore important to understand the influence of high-energy radiation on the chemistry in the envelopes around young stars in order to use molecules as tracers of X-rays and UV radiation. For this purpose we use detailed thermal balance and gas-phase chemical models to study the chemical evolution. In addition, we have started a JCMT program to search for chemical tracers such as CN, CO$^{+}$ and NO in the envelopes of both high- and low-mass stars, as well as intermediate mass YSOs.

\section{Observations and Results}
\label{sec:2}
\begin{figure}
\centering
\includegraphics[height=5.8cm,angle=-90]{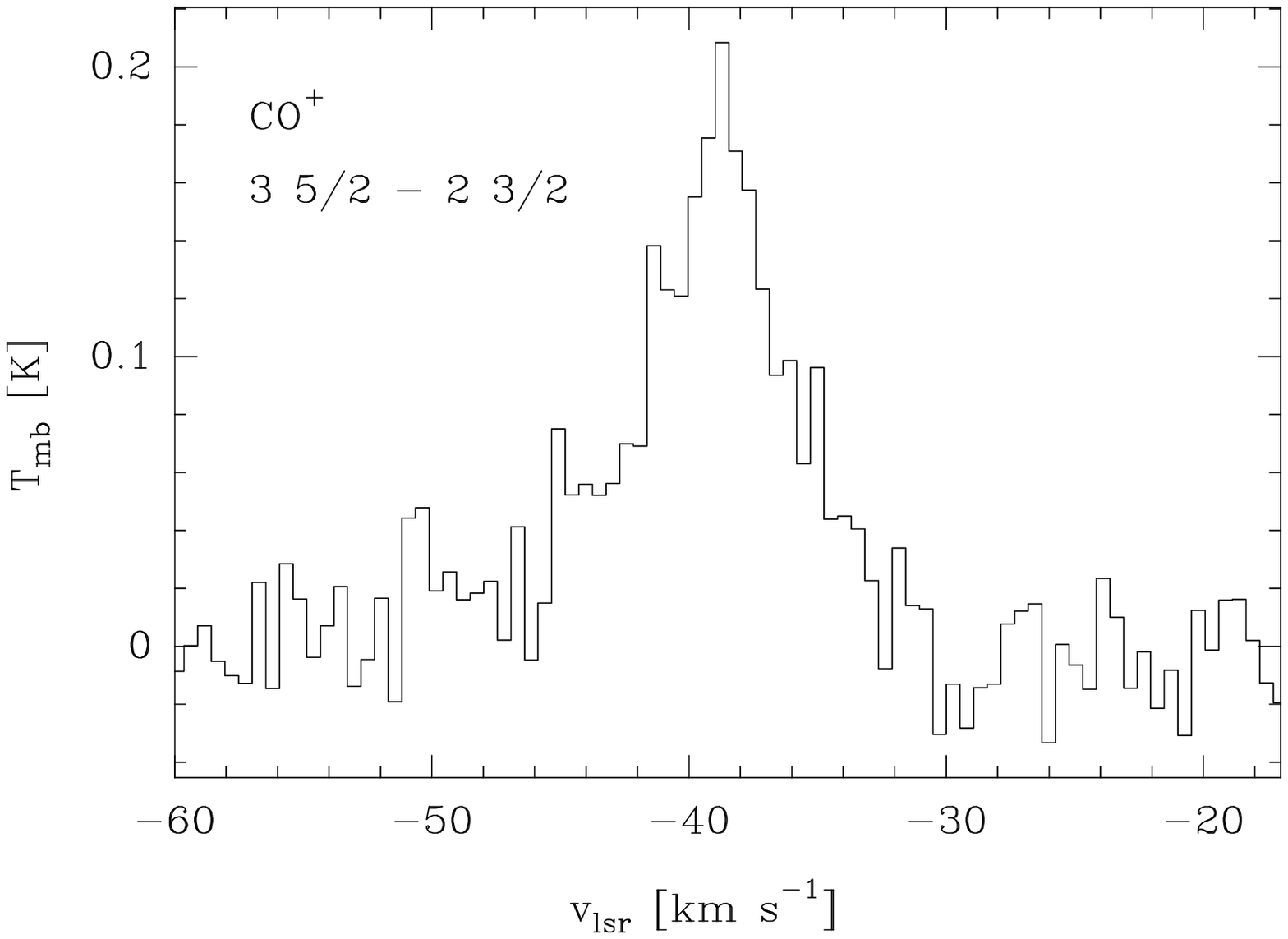}
\includegraphics[height=5.8cm,angle=-90]{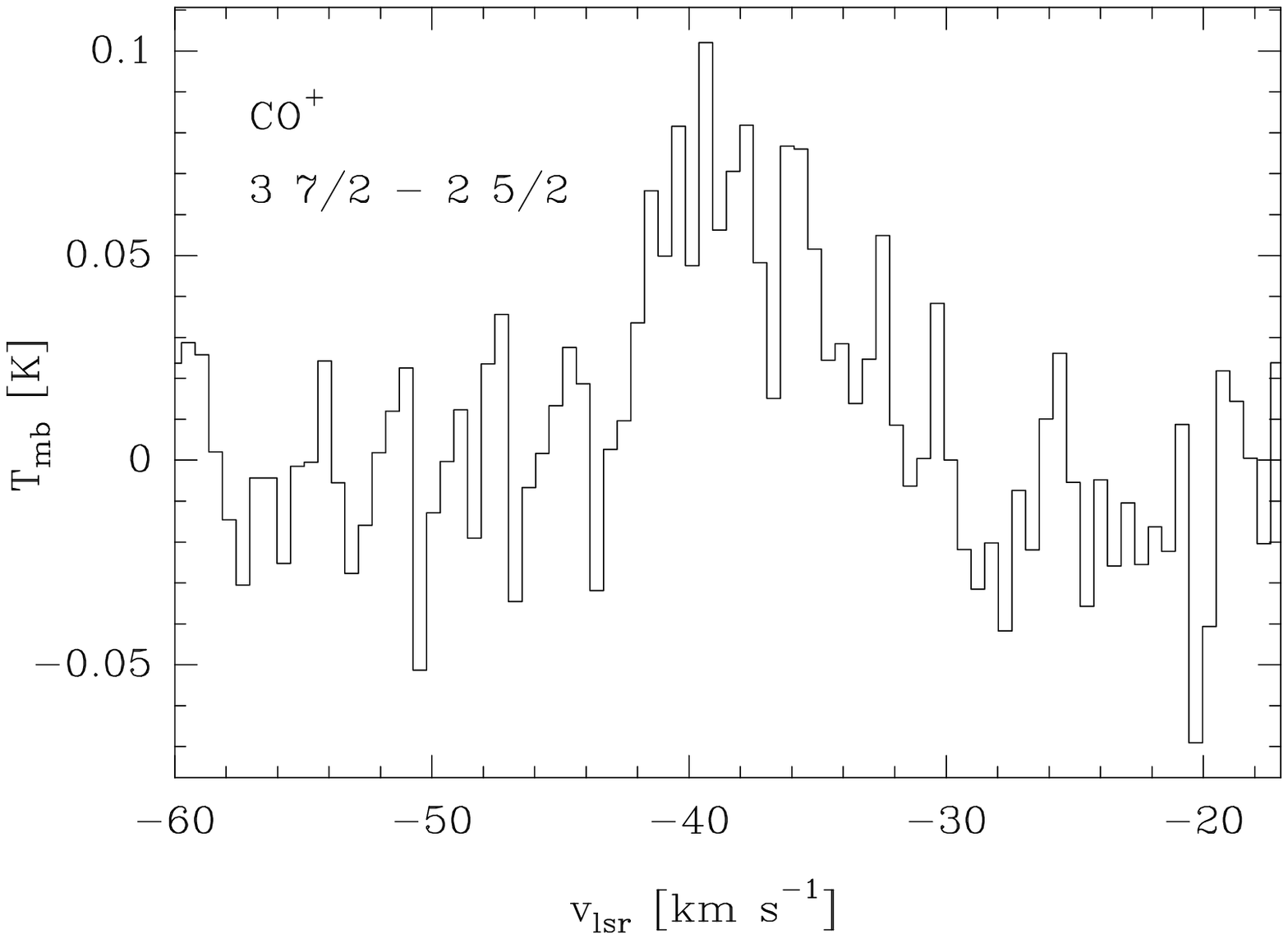}
\caption{\footnotesize CO$^+$ lines observed with the JCMT in 2003 toward W$3$ IRS$5$. The spectra have been calibrated in units of corrected main-beam brightness temperature T$_{mb}$.}
\label{pobs}       
\end{figure}

Hyperfine lines of CN and NO with high critical densities have been detected toward several sources and CO$^+$ toward W$3$ IRS$5$ with the $15$\,m James Clerk Maxwell Telescope (JCMT)\footnote{The James Clerk Maxwell Telescope is operated by the Joint Astronomy Centre, on behalf of the Particle Physics and Astronomy Research Council of the United Kingdom, the Netherlands Organization for Scientific Research, and the National Research Council of Canada.} on Mauna Kea, Hawaii during several runs in $2003$. For our observations we used the dual polarization receiver B$3$ at $345$\,GHz as the front end in order to probe the highest density material close to the protostars rather than the lower-density surrounding cloud material. Figure \ref{pobs} shows the observations of CO$^+$ and Table \ref{tobs} summarizes the results for the lines detected toward the massive star forming region W$3$ IRS$5$. The spectra have been reduced by using the GILDAS package. 

\begin{threeparttable}
\centering
\caption{Gaussian Fits to Observed Spectral Lines Toward W$3$ IRS$5$}
\label{tobs}       
\begin{tabular}{lccrcc}
\hline\noalign{\smallskip}
Molecule & Transition & Frequency & $\int T_{mb} dV$  & $\triangle$ V     & T$_{mb}$   \\
         &            &  [MHz]    & [K\,km\,s$^{-1}$] & [km\,s$^{-1}$] &     [K]        \\ 
\noalign{\smallskip}\hline\noalign{\smallskip}
CN\tnote{a}     & $3\frac{5}{2}\frac{3/5}{2} \rightarrow 2\frac{3}{2}\frac{1/3}{2}$ & 340035.4 & 15.30 & 2.80 & 2.48 \\
CN\tnote{a}     & $3\frac{5}{2}\frac{7}{2} \rightarrow 2\frac{3}{2}\frac{5}{2}$ & 340031.5 & 15.30 & 2.80 & 2.68 \\
CN\tnote{a}     & $3\frac{5}{2}\frac{3}{2} \rightarrow 2\frac{3}{2}\frac{3}{2}$ & 340019.6 & 15.30 & 2.80 & 0.58 \\
CN\tnote{a}     & $3\frac{5}{2}\frac{5}{2} \rightarrow 2\frac{3}{2}\frac{5}{2}$ & 340008.1 & 15.30 & 2.80 & 0.65 \\
CN\tnote{a}     & $3\frac{7}{2}\frac{7/9}{2} \rightarrow 2\frac{5}{2}\frac{5/7}{2}$ & 340247.8 & 20.80 & 4.20 & 4.70 \\
CN\tnote{a}     & $3\frac{7}{2}\frac{5}{2} \rightarrow 2\frac{5}{2}\frac{5}{2}$ & 340261.8 & 20.80 & 4.20 & 0.63 \\
CN\tnote{a}     & $3\frac{7}{2}\frac{7}{2} \rightarrow 2\frac{5}{2}\frac{7}{2}$ & 340265.0 & 20.80 & 4.20 & 0.63 \\
CO$^+$     & $3\frac{5}{2} \rightarrow 2\frac{3}{2}$ & 353741.3 & 0.98 & 4.70 & 0.20 \\
CO$^+$     & $3\frac{7}{2} \rightarrow 2\frac{5}{2}$ & 354014.2 & 0.43 & 4.70 & 0.09 \\
NO\tnote{a}     & $4\frac{7}{2}\frac{7}{2} \rightarrow 3\frac{5}{2}\frac{5}{2}$ & 351051.7 & 2.50 & 4.38 & 0.41 \\
NO\tnote{a}     & $4\frac{7}{2}\frac{9}{2} \rightarrow 3\frac{5}{2}\frac{7}{2}$ & 351043.5 & 2.50 & 4.38 & 0.33 \\
\noalign{\smallskip}\hline
\end{tabular}
\begin{tablenotes}
\item[a] \footnotesize Two or more hyperfine components have been fitted with the HFS fit method of CLASS. The line integral is the sum of the hyperfine components.
\end{tablenotes}
\end{threeparttable}
 
\section{Chemical Modeling}
To study the influence of X-rays and UV radiation from the young stellar object on its own environment, we used and developed the chemical models by Doty et al. ($2002$). The chemical model is based upon the UMIST gas-phase chemical reaction network (Millar et al. $1997$) and calculates the evolution of the cloud as a function of position and time. The density distribution and the temperature structure are taken from continuum observations of the dust and it is assumed that they do not change significantly with time. Given a chemical model, the millimeter line radiative transfer can be computed and convolved with the beam of the telescope. The observed lines can then be compared to the modeled lines.

\begin{figure}
\centering
\includegraphics[height=5.5cm,angle=-90]{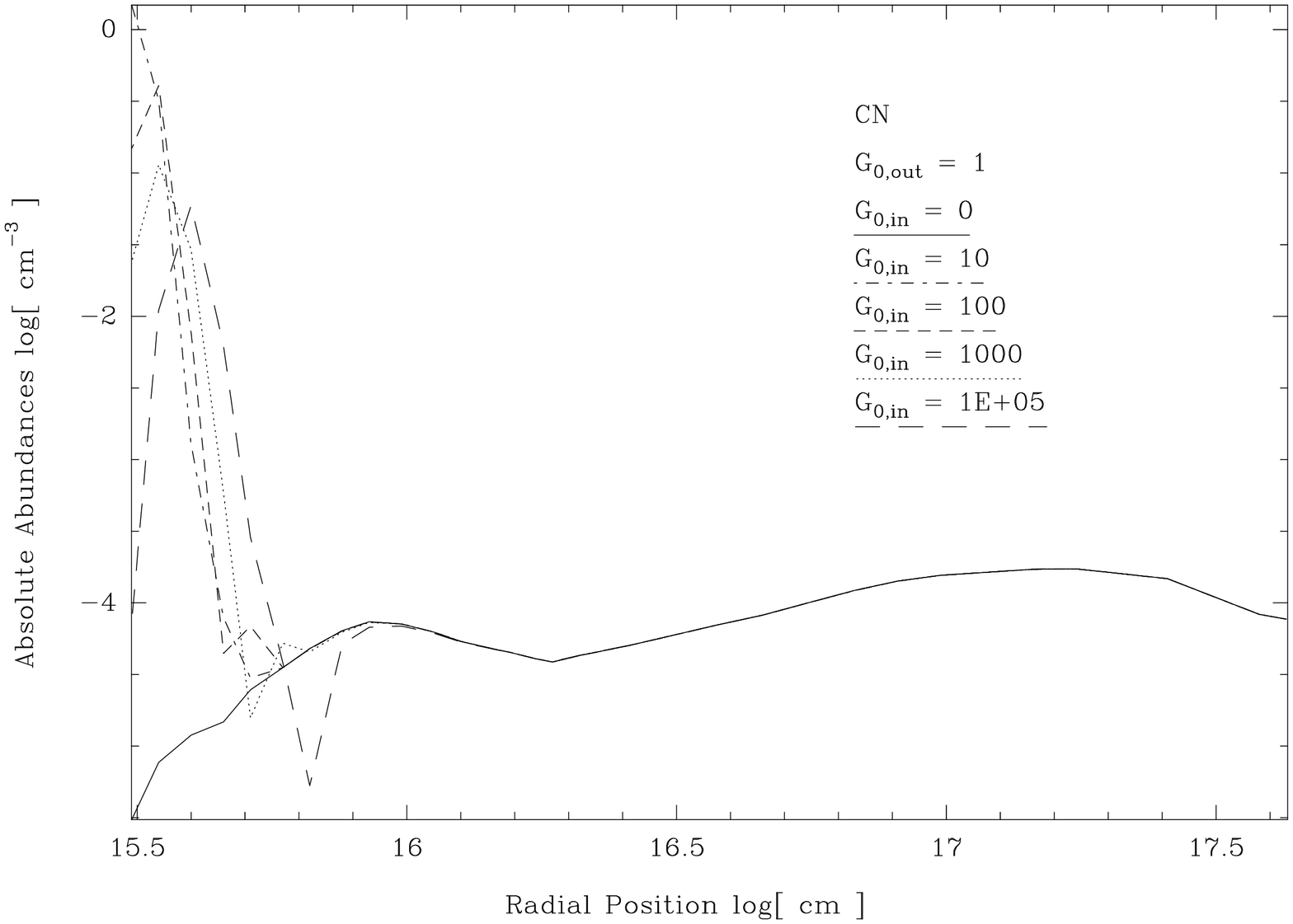}
\includegraphics[height=5.5cm,angle=-90]{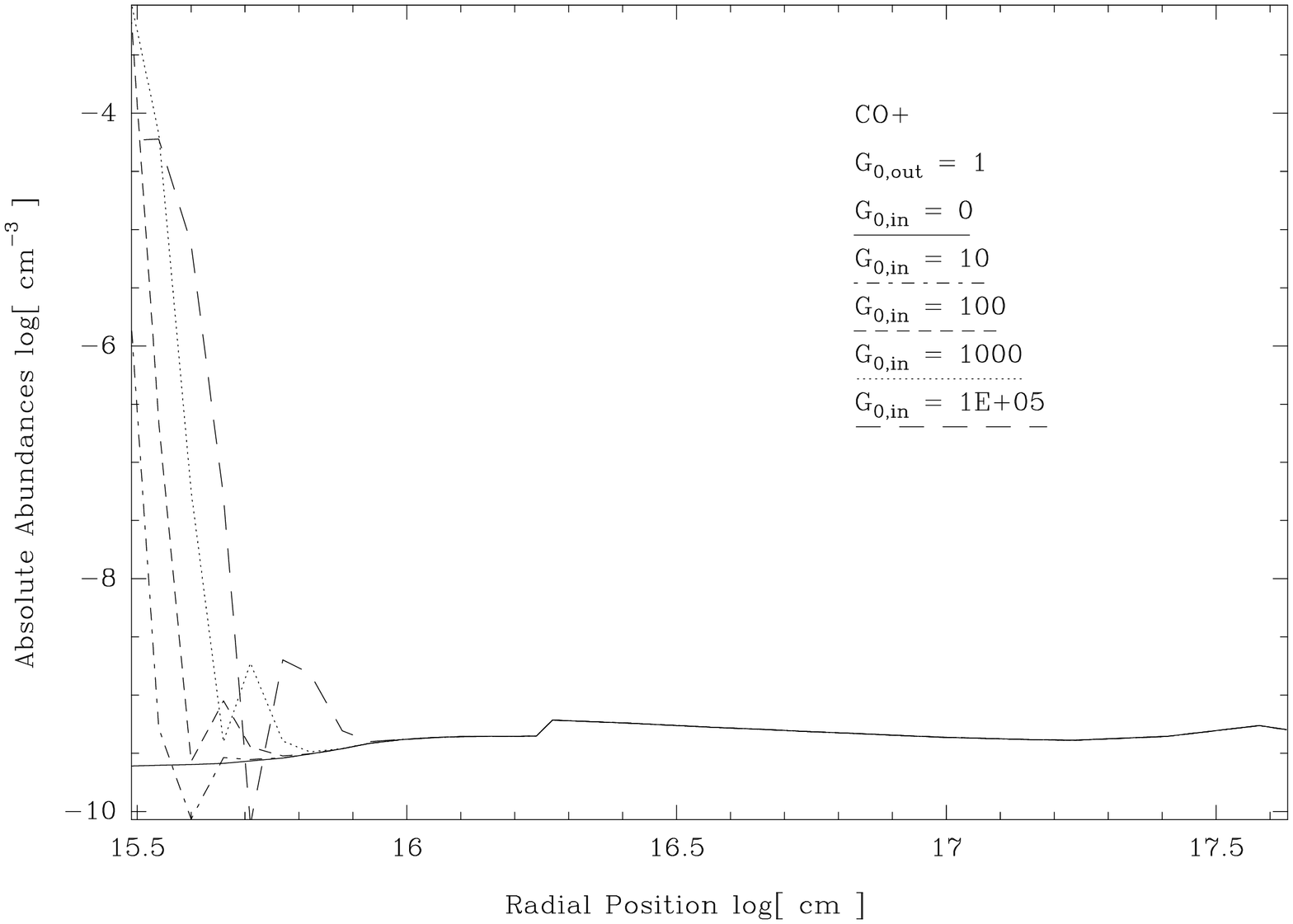}
\caption{\footnotesize Modeled CN and CO$^+$ abundances from a massive protostellar envelope assuming different UV radiation fields incident on the inner envelope. The UV flux at $10^{15.5}$\,cm is characterized in units of G$_{0,in}$, the enhancement of flux with respect to the average interstellar flux at $6<h\nu<13.6$\,eV of $1.6 \times 10^{-3}$\,erg\,cm$^{-2}$\,s$^{-1}$.}
\label{mod}
\end{figure}

As a first step, we have investigated the effects of UV radiation on the inner envelope (St\"auber et al., in preparation). Figure \ref{mod} shows that molecules like CN and CO$^+$ are clearly enhanced by orders of magnitude in the inner $600$\,AU. Assuming that the W$3$ IRS$5$ envelope is similar to that of AFGL $2591$, the observed CN can be reproduced quantitatively with a modest UV field. The intensity of CO$^+$, however, is clearly underestimated by the UV models. The most efficient reactions to build up CO$^+$ in the presence of UV photons from the central source are C$^+$ $+$ OH $\rightarrow$ CO$^+$ $+$ H and C$^+$ $+$ O$_2$ $\rightarrow$ CO$^+$ $+$ O, thus a lot of CO$^+$ is produced by C$^+$. Kramer et al. (in preparation) measured C$^+$ lines toward W$3$ IRS$5$ and modeled the PDR of the whole region with an appropriate UV field. The high C$^+$ abundances could not be modeled, however. This suggests that CO$^+$ might be a tracer for X-rays, which has also been suggested by other authors (e.g. Krolik \& Kallman $1983$). X-ray photons can ionize heavy elements like carbon very efficiently leading to multiply ionized species, which then recombine with electrons or react with H, He and H$_2$ via charge transfer very quickly to singly ionized species. In addition CO may be ionized by secondary electrons from primary X-ray ionizations of heavy elements. X-rays can also penetrate deeper into the cloud than UV photons due to smaller cross sections. It is therefore essential to investigate also the effects of X-rays on the chemistry in the envelopes of YSOs, what is planned to do as a next step. Such model results may also be important to predict chemical radiation probes for future instruments like Herschel, which will be able to observe hydrates like CH, CH$^+$, OH and H$_2$O, which are thought to be sensitive to high energy radiation.

\printindex
\end{document}